\documentclass[conference]{IEEEtran}
\IEEEoverridecommandlockouts
\usepackage[T1]{fontenc}
\usepackage[utf8]{inputenc}
\usepackage{amsmath,amssymb,amsfonts}
\usepackage{algorithmic}
\usepackage{graphicx}
\usepackage{textcomp}
\usepackage{xcolor}
\usepackage{longtable}
\usepackage{float}
\usepackage{makecell}
\usepackage{multirow}
\RequirePackage[            
    backend=biber,
    style=ieee
]{biblatex}
\def\BibTeX{{\rm B\kern-.05em{\sc i\kern-.025em b}\kern-.08em
    T\kern-.1667em\lower.7ex\hbox{E}\kern-.125emX}}
\begin{document}

\title{Improving the performance of an ASV system using hybrid speech features}

\author{\IEEEauthorblockN{1\textsuperscript{st} Stanisław Ciszkiewicz}
\IEEEauthorblockA{\textit{Warsaw University of Technology} \\
Warsaw, Poland \\
stanislaw.ciszkiewicz.stud@pw.edu.pl}
\and
\IEEEauthorblockN{2\textsuperscript{nd} Artur Janicki}
\IEEEauthorblockA{\textit{Warsaw University of Technology} \\
Warsaw, Poland \\
artur.janicki@pw.edu.pl}
}

\maketitle

\begin{abstract}
The growing need for secure and convenient authentication methods has led to the increasing popularity of biometric solutions. In addition to traditional and popular methods, such as fingerprint or iris scanning, voice-based approaches are also employed. User identity verification based on voice is conducted using Automatic Speaker Verification (ASV) systems. Despite their many advantages, these systems are sensitive to various types of attacks and acoustic noises, which can reduce verification accuracy. This work examines the potential to improve the performance of ASV systems by using hybrid feature sets that combine different signal representations, starting with widely-used Mel-Frequency Cepstral Coefficients (MFCC), through Constant Q Cepstral Coefficients (CQCC) and ending with the innovative RAB descriptor. Experiments were conducted on recordings from the Google Speech Commands dataset under two scenarios: in clean conditions and in the presence of acoustic noise. Finally, the systems' performance was compared using the EER metric to determine whether hybrid feature sets decrease verification error. The results show that using a hybrid feature set (PNCC+RAB) improves speaker verification performance under noisy conditions.
\end{abstract}


\section{Introduction}
Voice biometrics is a technology that enables the authentication and identification of a user based on their voice. Similar to other biometric methods, such as fingerprinting and iris scanning, it uses a person's unique biological and behavioral characteristics. Unlike traditional authentication methods based on something the user knows (e.g., a password or PIN) or something they possess (e.g., a card, token, or phone), biometric data is harder to break using brute-force or social engineering methods, while also eliminating the risk of forgetting a password or losing an authentication device. 

Despite the mentioned advantages, this technology may pose certain risks. The rapid development of speech synthesis tools and replay attacks~\cite{replay} raises justified concerns about the possibility of user impersonation and the compromise of authentication system integrity. Moreover, challenging acoustic conditions~\cite{noise_best}, such as background noise or channel distortions, may also affect the performance of voice biometric systems. 

In the literature, various feature extraction techniques can be found, ranging from traditional and widely used Mel-Frequency Cepstral Coefficients (MFCC)~\cite{mfcc}, which provide a good representation of speaker characteristics, through Constant Q Cepstral Coefficients (CQCC)~\cite{cqcc-info}, known for their robustness to replay attacks, to Power-Normalized Cepstral Coefficients (PNCC)~\cite{kim2}, designed to improve robustness to noise and speech signal distortions. Additionally, novel feature extraction approaches appear, such as the RAB descriptor based on a nonlinear model of the vocal folds~\cite{bronakowski}.

Despite the development of emerging feature extraction techniques, many methods still remain vulnerable to certain types of threats -- relying on a single feature type may reduce performance in certain scenarios. To enhance robustness across various conditions and threats, combining multiple features into a hybrid set is recommended~\cite{hybrid}, with the key challenge being selecting complementary features that collectively improve overall system performance.

The aim of this work is to improve the performance of an Automatic Speaker Verification (ASV) system by developing and implementing a hybrid set of acoustic features. The effectiveness of the proposed feature set will be evaluated on a prepared experimental setup.

The experiments will assess the performance of the ASV system both under clean conditions and in the presence of typical acoustic noises. The conducted tests and analysis of the results will enable the determination of whether and to what extent the use of a hybrid feature set improves system performance under various acoustic conditions.

Our paper is structured as follows. First, in~Section~\ref{sec:sota} we briefly depict related work in the discussed field. Next, in~Section~\ref{sec:methodology} we present the methodology of our experiments. Their results are described and discussed in~Section~\ref{sec:results}. The paper ends in~Section~\ref{sec:conclusion}, with the conclusions. 
\section{Related work}
\label{sec:sota}

The ASV system is designed to determine whether the voice sample it receives actually belongs to the claimed user. 
The verification process is typically divided into two main stages~\cite{summary_asv}: 

\begin{itemize}
    \item \textbf{feature extraction} (frontend) -- responsible for extracting speaker-characteristic information from a voice sample,
    \item \textbf{classification} (backend) -- consists of models/methods that compare extracted features with the stored speaker model and perform the final verification decision.
\end{itemize}

\begin{figure}[H]
    \centering
    \includegraphics[width=\linewidth]{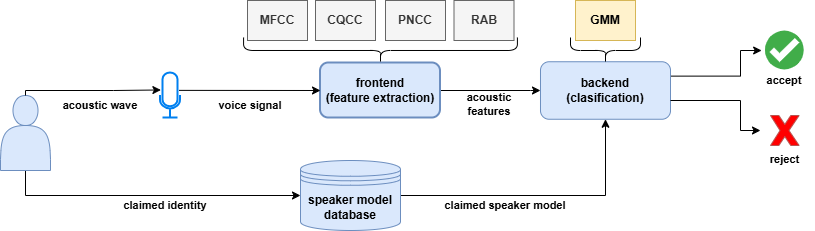}
    \caption{ASV system architecture used in experiments}
    \label{ASV_arch}
\end{figure}

The development of a secure and effective ASV system strongly depends on the choice of an appropriate feature extraction technique. The purpose of the extraction process is to transform a continuous audio signal into discrete acoustic feature vectors, which are then used as input for machine learning models or deep neural networks. Throughout the years, a wide range of feature extraction techniques has been proposed, each characterized by diverse applications within ASV systems, varying degrees of robustness to acoustic noise, and differing levels of vulnerability to potential attacks.

\subsection{Classical speech features used in ASV systems}

MFCC features are the most traditional and commonly used acoustic features in ASV systems. They are mel-cepstral features that describe the overall shape of the signal spectrum after it has been transformed to the mel scale, reflecting the nonlinear characteristics of human hearing. 

The MFCC extraction process~\cite{mfcc} consists of several operations, such as pre-emphasis, framing, windowing, and discrete Fourier transform (DFT). The obtained spectrum is then passed through a mel filter bank, which maps it onto the nonlinear mel scale. Finally, the discrete cosine transform (DCT) is applied to produce the final set of MFCC coefficients.

CQCC features~\cite{cqcc-info} are an alternative to traditional MFCC features. The extraction process is similar to MFCC, with one significant difference -- in CQCC, instead of the DFT, the Constant-Q transform is used. The use of the Constant-Q transform provides variable resolution, offering higher frequency resolution at low frequencies and higher temporal resolution at high frequencies, which enables more precise representation of the speech signal. It is highlighted that CQCC features are particularly effective in ASV systems, especially in improving robustness against replay attacks~\cite{cqcc-replay}, making them suitable for spoofing detection tasks in ASV systems.

\begin{figure}[H]
    \centering
    \includegraphics[width=0.95\linewidth]{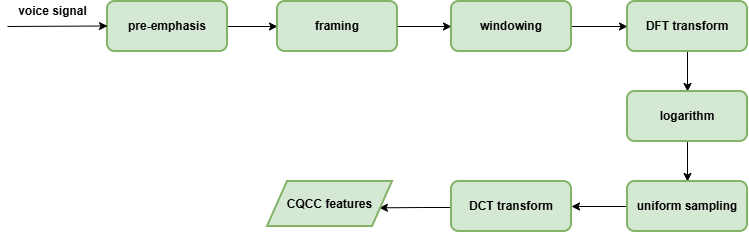}
    \caption{CQCC features extraction schema}
    \label{cqcc}
\end{figure}

It has been observed that MFCC performance significantly degrades in noisy conditions. To address this issue, Kim and Stern proposed Power-Normalized Cepstral Coefficients (PNCC)~\cite{kim2}, designed to improve the robustness of ASV systems to acoustic noise. While its initial steps are similar to MFCC, PNCC introduces key modifications: a Gammatone filterbank that models human auditory frequency perception more accurately than the Mel filterbank, and a power-law nonlinearity rather than logarithmic compression, preserving important speech components under noisy conditions.

\subsection{RAB descriptor}
Apart from cepstral features, which are based on the shape of the signal spectrum, many scientists have proposed alternative feature extraction techniques, e.g., Log Area Ratio (LAR) coefficients~\cite{LAR}.
The author of~\cite{bronakowski} proposed the use of the so-called RAB descriptor, based on a simple nonlinear model of the vocal folds~\cite{rab2}.

The model selected by the author can be described by the following equation:
\begin{equation}\label{eq:rabc}
s(x, \dot{x}) = kx + a x \dot{x},
\end{equation}
where $k$ determines the stiffness of the vocal fold tissue, whereas $a$
controls the degree of vocal fold tension.

\begin{figure}[H]
    \centering
    \includegraphics[width=0.95\linewidth]{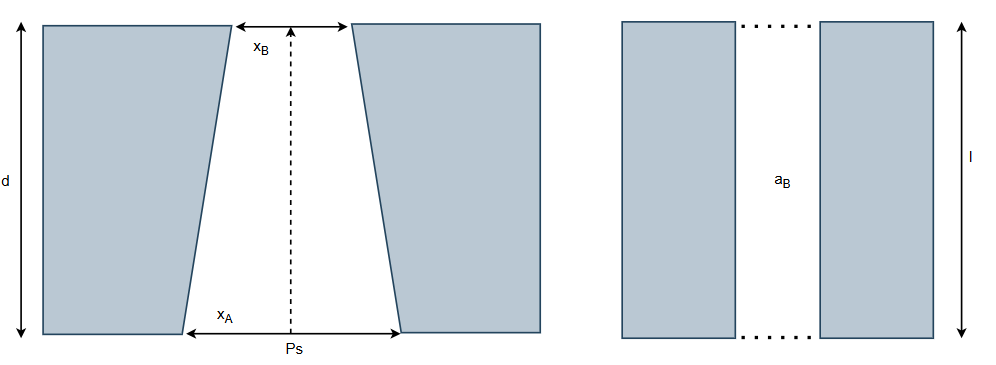}
    \caption{Simple schema of a nonlinear model of the vocal folds (based on~\cite{rab2})}
    \label{rabfolds}
\end{figure}

The nonlinear model equation can also be expressed in a discrete form:

{\scriptsize
\begin{equation} \label{eq:dyskretny}
m \frac{x_{n+1} - 2x_n + x_{n-1}}{\Delta t^2} 
+ r \frac{x_n - x_{n-1}}{\Delta t} 
+ a x_n \frac{x_n - x_{n-1}}{\Delta t} 
- b + k x_n + c \, \Theta(x_n) x_n^2 = 0
\end{equation}
}

By transforming the discrete relationship as in Equation~\eqref{eq:dyskretny}, it is possible to determine five key physical parameters of the model:

{\footnotesize
\begin{equation} \label{eq:RABKC}
R = \frac{r \, \Delta t}{m}, \quad
A = \frac{a \, \Delta t}{m}, \quad
B = \frac{b \, \Delta t^2}{m}, \quad
K = \frac{k \, \Delta t^2}{m}, \quad
C = \frac{c \, \Delta t^2}{m}
\end{equation}
}

After substituting the above relationships for the parameters \eqref{eq:RABKC} into Equation~\eqref{eq:dyskretny} and performing the transformation, the final form of the equation for the sample $x_{n+1}$ is obtained.

{\scriptsize
\begin{equation} \label{eq:xnp1}
x_{n+1} = 2x_n - x_{n-1} - R(x_n - x_{n-1}) - A x_n (x_n - x_{n-1}) + B - K x_n - C \, \Theta(x_n) x_n^2
\end{equation}
}

By moving the terms independent of the coefficients $R$, $A$, $B$, $K$, and $C$ to the left-hand side of the equation, we obtain:

{\footnotesize
\begin{equation} \label{eq:ostateczne_równanie}
d_{n}= 
R (x_n - x_{n-1}) 
+ A \, x_n (x_n - x_{n-1}) 
- B + K \, x_n + C \ \Theta(x_n) x_n^2, 
\end{equation}
}
where
{\footnotesize
\begin{equation} \label{eq:dn}
d_n = -(x_{n+1} - 2 x_n + x_{n-1})
\end{equation}
}

The presented Equation \eqref{eq:ostateczne_równanie} can be expressed in matrix form:
{\footnotesize
\begin{equation}\label{eq:matrix1}
{d} = {M} \, {a}
\end{equation}
}

The matrix Equation \eqref{eq:matrix1} represents the regression model in terms of the parameters $R$, $A$, $B$, $K$, and $C$. These parameters can be estimated using the least squares method.

Experiments conducted in~\cite{bronakowski} revealed that the variance of two of them, $K$ and $C$, is low across all speakers. As a result, they were removed from the final descriptor, which finally consists of:

{\footnotesize
\begin{equation}
\mathbf{a} = [R, A, B]^\top
\end{equation}
}
\noindent and, hereinafter, is referred to as the RAB descriptor.

\section{Methodology}
\label{sec:methodology}

\subsection{Analyzed feature extraction techniques}

The feature extraction techniques selected for the experiments include MFCC, CQCC, PNCC, and the RAB descriptor. This choice was motivated by the fact that each of them has a diverse range of applications and provides robustness to different types of distortions. For this reason, in addition to evaluating the features individually, their combinations forming hybrid feature sets were also investigated.

All feature vectors were normalized, with normalization parameters estimated from the training data and applied consistently across all experiments.
\subsection{Speaker modeling}
Gaussian Mixture Models (GMMs)~\cite{gmm} were chosen because they are the most classical and widely used approach in ASV systems, especially in constrained resource settings. More advanced, newer methods, such as hybrid neural network architectures like ECAPA2~\cite{ecapa2} or CAM++~\cite{cam++}, were not considered in order to isolate the impact of feature extraction techniques within a fixed classification backend. GMMs assume that the data can be modeled as a mixture of a finite number of Gaussian distributions (16 in this work), whose combination permits approximating the overall distribution of the input features. 

In the proposed system, a Universal Background Model (UBM)~\cite{gmm-ubm} was created to present general speaker characteristics. In the next step, individual speaker models were obtained from the UBM using MAP adaptation~\cite{REYNOLDS-map}.

\subsection{Dataset}
The experimental ASV system was based on the Google Speech Commands (GSC) dataset~\cite{gsc_art}, a widely used corpus in speaker recognition and verification tasks~\cite{GSCusage,gscasv2}. Although datasets such as VoxCeleb~\cite{voxceleb} are popular in ASV-related studies due to their challenging real-world recording conditions, they are mainly designed for text-independent speaker verification. Since the objective of this study was to investigate text-dependent speaker verification, the GSC dataset was adopted. 

The dataset contains approximately 65,000 short recordings in English consisting of simple words such as ``bird'', ``bed'', and ``six''. Due to the risk of underrepresentation (not every speaker has the same number of recordings), 30 speakers with the most recordings were selected. For each of them, 100 recordings were randomly selected and split into training and testing sets in a 4:1 ratio. The preliminary analysis revealed that individual recordings (lasting 1-2 seconds) were too short, and using them led to unstable results. For this reason, they were concatenated into longer sequences of five recordings. Finally, for each speaker, 16 training recordings and 4 test recordings were obtained.

A key aspect is understanding the methodology used for testing the developed ASV system. To verify all hypothetical situations that can occur in the ASV system, each of the 120 recordings (30 speakers × 4 recordings) was compared against each of the 30 models. In total, for each setup, 3,600 tests (120 recordings × 30 models) were conducted, enabling a thorough assessment of the system’s performance across different scenarios -- both when a sample comes from the correct speaker and when it comes from an unauthorized individual.

\subsection{Acoustic conditions}

\subsubsection{Clean conditions.}
In the clean condition setup, all recordings were evaluated without additional background noise. This scenario represents ideal operating conditions for the ASV system, enabling assessment of its baseline performance when speech signals are clear and undistorted.

\subsubsection{Noisy conditions.}
In the noisy condition setup, the recordings were corrupted with additive background noise to simulate more realistic environments. To simulate the noisy scenario, the NoiseX-92 database~\cite{noisex92} was used. The conducted experiments focused on two types of acoustic noise: babble noise (conversations of multiple people in offices or cafés) and volvo noise (car engine noise). This evaluation enables analysis of the system’s robustness and performance degradation under adverse acoustic conditions.

The proposed experimental setup consisted of training the models on clean recordings, followed by testing on recordings with added noise. To evaluate the systems' performance across various noise levels, test recordings were prepared at signal-to-noise ratios (SNRs) of 0, 5, 10, 15, and 20~dB.

\section{Results}
\label{sec:results}
The experiments focused on evaluating the potential to improve ASV system performance by combining acoustic features into hybrid feature sets. The effectiveness of various feature sets was examined in both clean and noisy conditions.

\subsection{Clean conditions}
In the first scenario, with training and testing data free from additional acoustic distortions, experiments were conducted on both individual and hybrid feature sets. The EER values presented in Table~\ref{tab:eer_clean} indicate that hybrid feature sets mostly performed similarly to or better than individual acoustic features. However, it should be noted that under clean conditions, the systems based on MFCC or PNCC achieved almost perfect performance (EERs of $0.11\%$ and $0.10\%$, respectively), leaving not much room for further improvement. 

\begin{table}[H]
\caption{EER [\%] for various feature sets in clean conditions}
\begin{center}
\footnotesize
\begin{tabular}{|c|c|}
\hline
\textbf{Feature Set} & \textbf{EER [\%]} \\
\hline
CQCC & 8.48 \\
\hline
MFCC & 0.11 \\
\hline
PNCC & 0.10 \\
\hline
RAB & 10.79 \\
\hline
CQCC + RAB & 6.90 \\
\hline
CQCC + PNCC & 0.85 \\
\hline
CQCC + PNCC + RAB & 0.19 \\
\hline
CQCC + MFCC & 0.98 \\
\hline
MFCC + PNCC + RAB & \textbf{0.06} \\
\hline
PNCC + RAB & 0.13 \\
\hline
\end{tabular}
\label{tab:eer_clean}
\end{center}
\normalsize
\end{table}

\subsection{Noisy conditions}

Table~\ref{tab:eer_babble} presents the EER values obtained in the experiments with babble noise. Examining the results, it can be observed that the ASV system performs worst when using RAB features, while the best performance is achieved with feature sets containing PNCC, particularly PNCC and PNCC+RAB. It should be emphasized that adding three RAB coefficients to the PNCC features results in a decrease of EER of up to 4 percentage points (for an SNR of 0 dB). For MFCC features, a noticeable drop in performance was observed compared to the first scenario. Under clean conditions, they achieved an EER of $0.11\%$ (placing them at the top of the ranking), whereas in babble noise at an SNR of 0 dB, their EER was over 13 percentage points higher than that of PNCC+RAB features.

\setlength{\tabcolsep}{7pt}
\begin{table}[H]
\caption{EER [\%] for various feature sets -- babble noise}
\begin{center}
\footnotesize
\begin{tabular}{|c|c|c|c|c|c|}
\hline
\multirow{2}{*}{\makecell[c]{\textbf{Feature Set}}} & \multicolumn{5}{c|}{\textbf{SNR [dB]}} \\ \cline{2-6}
 & 0 & 5 & 10 & 15 & 20 \\ \hline
CQCC & 46.74 & 41.54 & 37.33 & 31.74 & 25.82 \\
\hline
MFCC & 42.51 & 39.28 & 32.53 & 26.71 & 20.85 \\
\hline
PNCC & 33.19 & 25.03 & \textbf{19.04} & 14.73 & 10.06 \\
\hline
RAB & 45.01 & 44.96 & 40.91 & 38.99 & 37.43 \\
\hline
CQCC + RAB & 42.47 & 39.27 & 33.19 & 28.99 & 25.34 \\
\hline
CQCC + MFCC & 42.60 & 35.99 & 30.85 & 28.39 & 25.78 \\
\hline
CQCC + PNCC & 37.61 & 32.51 & 27.67 & 20.91 & 16.84 \\
\hline
CQCC + PNCC + RAB & 34.77 & 29.22 & 22.46 & 16.55 & 12.46 \\
\hline
MFCC + PNCC + RAB & 37.46 & 31.47 & 25.78 & 17.49 & 11.87 \\
\hline
PNCC + RAB & \textbf{29.17} & \textbf{24.83} & 20.07 & \textbf{11.54} & \textbf{7.47} \\
\hline
\end{tabular}
\label{tab:eer_babble}
\end{center}
\normalsize
\end{table}

\begin{figure}[H]
    \centering \includegraphics[width=1\linewidth]{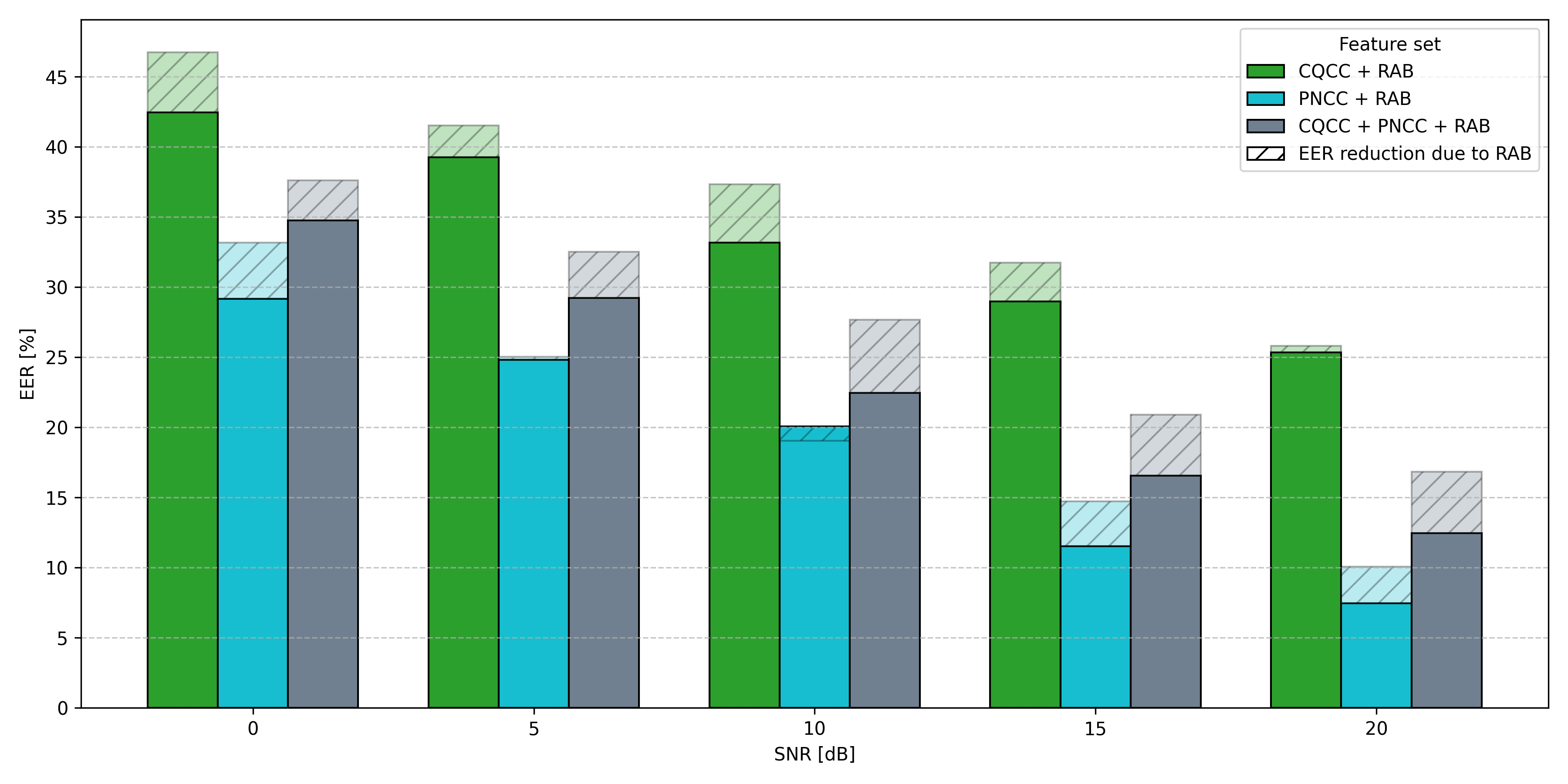}
    	\caption{Impact of RAB on EER across SNR levels for various feature sets -- babble noise}
    \label{fig:eer_babble}
\end{figure}

Figure~\ref{fig:eer_babble} shows the dependence of EER on SNR for various feature sets, with and without RAB coefficients under babble noise. A clear, consistent reduction in EER is observed for feature sets including RAB coefficients. Only in a few cases (with PNCC for 5 dB and with CQCC for 20 dB) this reduction was negligible.

The results of the experiments with Volvo noise are presented in Table \ref{tab:eer_volvo}. It can again be observed that the best EER values are achieved by systems based on feature sets that include PNCC. This includes both the individual PNCC features and the hybrid set PNCC+CQCC+RAB. As in the babble noise scenario, the best performance was observed for the PNCC+RAB set.

\setlength{\tabcolsep}{7pt}
\begin{table}[H]
\caption{EER [\%] for various feature sets -- volvo noise}
\begin{center}
\footnotesize
\begin{tabular}{|c|c|c|c|c|c|}
\hline
\multirow{2}{*}{\makecell[c]{\textbf{Feature set}}} & \multicolumn{5}{c|}{\textbf{SNR [dB]}} \\ \cline{2-6}
 & 0 & 5 & 10 & 15 & 20 \\ \hline
CQCC & 34.78 & 30.89 & 28.28 & 23.30 & 21.67 \\
\hline
MFCC & 33.53 & 31.77 & 29.30 & 25.01 & 20.09 \\
\hline
PNCC & 26.70 & 22.41 & 17.60 & 10.82 & 8.15 \\
\hline
RAB & 30.91 & 31.82 & 30.98 & 28.32 & 22.39 \\
\hline
CQCC + MFCC & 33.26 & 32.69 & 33.79 & 32.54 & 30.09 \\
\hline
CQCC + PNCC & 29.12 & 24.17 & 18.15 & 10.80 & 6.59 \\
\hline
CQCC + PNCC + RAB & 24.27 & 17.56 & \textbf{11.61} & 7.53 & 5.07 \\
\hline
CQCC + RAB & 31.81 & 30.68 & 25.69 & 22.50 & 21.21 \\
\hline
MFCC + PNCC + RAB & 27.40 & 23.22 & 17.61 & 14.04 & 11.74 \\
\hline
PNCC + RAB & \textbf{24.02} & \textbf{15.59} & 12.49 & \textbf{7.18} & \textbf{5.91} \\
\hline
\end{tabular}
\label{tab:eer_volvo}
\end{center}
\normalsize
\end{table}

\begin{figure}[H]
    \centering \includegraphics[width=1\linewidth]{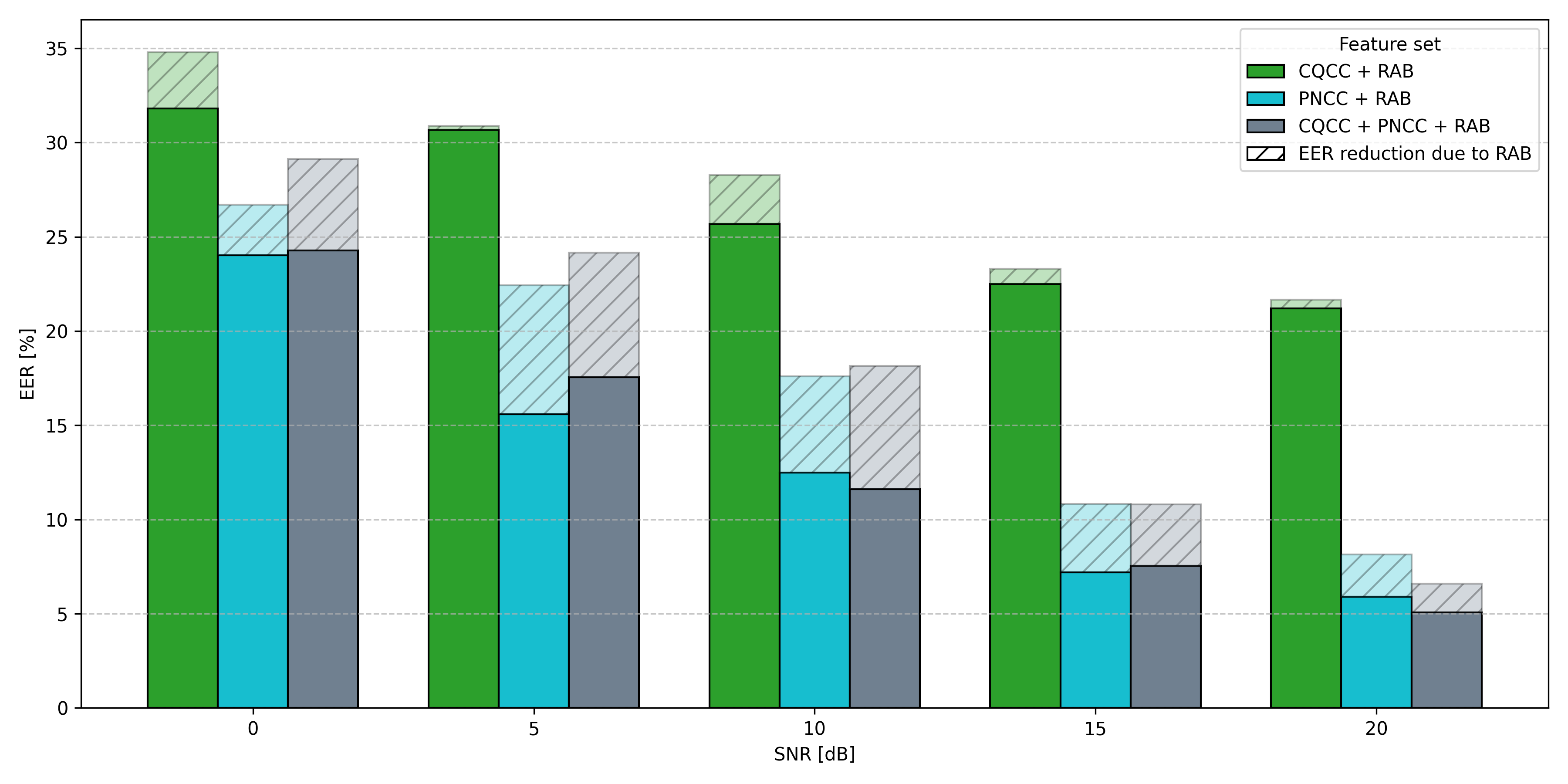}
    	\caption{Impact of RAB on EER across SNR levels for various feature sets -- Volvo noise}
    \label{fig:eer_volvo}
\end{figure}

Figure~\ref{fig:eer_volvo}, similarly to the babble noise scenario, illustrates a noticeable reduction in EER for feature sets containing RAB coefficients, which is especially well visible for feature sets with PNCC and CQCC+PNCC. 

\section{Conclusion}
\label{sec:conclusion}

This study examined the performance of an ASV system under both clean and noisy conditions, with a main focus on determining the effectiveness of both individual and hybrid feature sets. 

The conducted experiments showed that under clean conditions, using hybrid feature sets does not provide a significant benefit, as the results are comparable to those obtained using MFCC and PNCC features. In noisy conditions, PNCC-based feature sets clearly outperform the others, with the best results achieved by combining PNCC with the nonlinear RAB descriptor.

For both evaluated noise types, the hybrid PNCC+RAB feature set reduced EER compared to PNCC alone by up to 4 percentage points in babble noise and up to 7 percentage points in Volvo noise.

Although the results of adding the RAB descriptor to other features, especially PNCC and MFCC, are highly encouraging, they should be confirmed in a larger study with more speakers. Future work can also involve using other speaker verification methods, in particular, neural ones, such as the previously mentioned ECAPA2 or CAM++.

\printbibliography
\end{document}